\documentclass[conference]{IEEEtran}
\IEEEoverridecommandlockouts
\usepackage{textcomp}
\usepackage[nocompress]{cite}

\usepackage{epsfig, epstopdf, graphicx, balance}
\usepackage{amsmath,amssymb, mathrsfs, amsfonts}
\usepackage[mathscr]{euscript}
\usepackage{microtype} 
\usepackage{tabulary} 

\usepackage{enumitem}
\setlist[itemize]{itemsep=3pt,topsep=0pt,parsep=0pt,partopsep=3pt,leftmargin=2em}
\setlist[enumerate]{itemsep=3pt,topsep=0pt,parsep=0pt,partopsep=3pt,leftmargin=2em}

\usepackage{hyperref}

\usepackage[scaled=0.85]{beramono}

\usepackage{amsthm}
\theoremstyle{definition}
 
\theoremstyle{plain}

\usepackage{mathtools}

\usepackage[linesnumbered,ruled,vlined]{algorithm2e}
\SetAlgorithmName{Algorithm}{Algorithm}{Algorithm}
\IncMargin{0.5em}
\SetCommentSty{textnormal}
\SetNlSty{}{}{:}
\SetAlgoNlRelativeSize{0}
\SetKwInput{KwGlobal}{Global}
\SetKwInput{KwPrecondition}{Precondition}
\SetKwProg{Proc}{Procedure}{:}{}
\SetKwProg{Func}{Function}{:}{}
\SetKw{And}{and}
\SetKw{Or}{or}
\SetKw{To}{to}
\SetKw{DownTo}{downto}
\SetKw{Break}{break}
\SetKw{Continue}{continue}
\SetKw{SuchThat}{\textit{s.t.}}
\SetKw{WithRespectTo}{\textit{wrt}}
\SetKw{Iff}{\textit{iff.}}
\SetKw{MaxOf}{\textit{max of}}
\SetKw{MinOf}{\textit{min of}}
\SetKwBlock{Match}{match}{}{}

\usepackage{array}
\newcolumntype{L}[1]{>{\raggedright\let\newline\\\arraybackslash\hspace{0pt}}m{#1}}
\newcolumntype{C}[1]{>{\centering\let\newline\\\arraybackslash\hspace{0pt}}m{#1}}
\newcolumntype{R}[1]{>{\raggedleft\let\newline\\\arraybackslash\hspace{0pt}}m{#1}}

\usepackage{color}
\usepackage[usenames,svgnames]{xcolor}
\usepackage{soul}
\soulregister\cite7
\soulregister\ref7
\soulregister\pageref7
\soulregister\autoref7
\soulregister\eqref7

\ifdefined\NoHighlight

\fi

\usepackage{multicol, multirow}

\usepackage{subfigure}

\renewcommand{\eqref}{Equation~\ref}

\newcommand*{\PicDir}{pic}
\newcommand*{\ExpDir}{exp}
\newcommand*{\Pic}[2]{\PicDir /#2.#1}
\newcommand*{\Exp}[2]{\ExpDir /#2.#1}

\newcommand*{\FigBottomVSpace}{\vspace*{-2ex}}

\newcommand*{\locec}{LoCEC}
\newcommand*{\loceccnn}{\locec{}-CNN}
\newcommand*{\locecxgb}{\locec{}-XGB}
\newcommand*{\commcnn}{CommCNN}
\newcommand*{\probwp}{ProbWP}
\newcommand*{\economix}{Economix}
\newcommand*{\xgb}{XGBoost}

\newcommand*{\wechat}{WeChat}
\newcommand*{\moments}{Moments}

\newcommand*{\ie}{\textit{i.e.}\@\xspace}
\newcommand*{\etal}{\textit{et. al.}\@\xspace}

\begin{document}

\title{\locec{}: Local Community-based Edge Classification in Large Online Social Networks}

\newcommand{\AuthNameSep}{\hspace*{1em}}  
\newcommand{\AuthNameVSep}{\vspace{.5ex}}  %
\newcommand{\AuthNameVSpace}{\vspace{1.2ex}}
\newcommand{\AuthAffSep}{\hspace{1.5em}}
\newcommand{\AuthAffVSep}{}  
\newcommand{\AuthAffVSpace}{\vspace{.8ex}}
\newcommand{\AuthEmailSep}{\hspace{2em}}
\newcommand{\AuthEmailVSep}{\vspace{.2ex}}  %

\newcommand*{\AuthName}[2]{#2$^{#1}$}
\newcommand*{\AuthNameWithNote}[3]{\AuthName{#1}{#2}\titlenote{#3}}
\newcommand*{\AuthAff}[2]{$^{#1}${#2}}
\newcommand*{\AuthEmail}[2]{$^{#1}${#2}}
\newcommand*{\AuthEmails}[3]{$^{#1}${\{#3\}@#2}}

\author{
	\IEEEauthorblockN{Chonggang Song$^{\dag}$, Qian Lin$^{\S}$, Guohui Ling$^{\dag}$, Zongyi Zhang$^{\dag}$, Hongzhao Chen$^{\dag}$, Jun Liao$^{\dag}$, Chuan Chen$^{\dag}$}
	\AuthNameVSpace
	\IEEEauthorblockA{
		$^{\dag}$Tencent Inc.\AuthAffSep
		$^{\S}$National University of Singapore, 
	}
	\AuthAffVSpace
	\IEEEauthorblockA{
		\AuthEmails{\dag}{tencent.com}{jerrycgsong, randyling, zongyizhang, tylerhzchen,  jensenliao,  chuanchen}\AuthEmailSep
		\AuthEmail{\S}{linqian@comp.nus.edu.sg}
	}
}

\maketitle

\begin{abstract}
Relationships in online social networks often imply social connections in real life.
An accurate understanding of relationship types benefits many applications, e.g.\ social advertising and recommendation. 
Some recent attempts have been proposed to classify user relationships into predefined types with the help of pre-labeled relationships or abundant interaction features on relationships.
Unfortunately, both relationship feature data and label data are very sparse in real social platforms like \wechat{}, rendering existing methods inapplicable.

In this paper, we present an in-depth analysis of \wechat{} relationships to identify the major challenges for the relationship classification task. 
To tackle the challenges, we propose a Local Community-based Edge Classification (\locec{}) framework that classifies user relationships in a social network into real-world social connection types. 
\locec{} enforces a three-phase processing, namely local community detection, community classification and relationship classification, to address the sparsity issue of relationship features and relationship labels. 
Moreover, \locec{} is designed to handle large-scale networks by allowing parallel and distributed processing.
We conduct extensive experiments on the real-world \wechat{} network with hundreds of billions of edges to validate the effectiveness and efficiency of \locec{}.
\end{abstract}

\begin{IEEEkeywords}
\wechat{}, edge classification, social networks
\end{IEEEkeywords}

\section{Introduction}

Understanding user relationship types benefits many social applications. 
An intuitive example would be social advertising~\cite{bakshy:ec12, aslay:wsdm18, aslay:vldb2017, wang:vldb2017, sigmodrec16:Wang}. 
When placing an advertisement for furniture and household items, it is best to select users that are family members with each other in order to boost online interaction as well as offline discussions of the product. 
Another example application is social recommendation~\cite{zhang:cikm16}. 
Users tend to have more interest in news articles that are commonly liked by their colleagues or games that are preferred by their schoolmates. 

Identifying the classes of edges in social networks has been extensively studied in literature. 
The works in \cite{leskovec:www10, agrawal:ijcai13, yang:sigir12} focus on classifying edges as positive or negative, representing friends or enemies.
Other works in \cite{wang:kdd10, backstrom:cscw14, li:www14} aim to identify a particular type of relationship such as advisor-advisee, employer-employee or romantic relationships.
Recent works in \cite{aggarwal:icde16, aggarwal:icde17} formally define the edge classification problem in general. 
A label propagation algorithm based on structural similarity is proposed in~\cite{aggarwal:icde16}, which presumes a large proportion of edge labels are available in order to propagate true class labels to the entire network. 
However, in a real-world social network, only a small percentage of edge labels could be obtained, rendering the above technique inadequate. 
The work in \cite{aggarwal:icde17} considers some edges are associated with textual content, such as communication history, and proposes a matrix-factorization algorithm to utilize both textual and structural information.
However, in real-life social platforms, users tend to communicate with only a small ratio of friends. 
For example, in \wechat{}, around $60\%$ of user pairs have no interactions over a month, making it difficult  to extract user-to-user features for analysis. 
Furthermore, it is hard for the approach in \cite{aggarwal:icde17} to process a network with hundreds of billions of edges since it requires to construct matrices whose columns represent all edges in the network.


In this work, we first carefully analyze how different types of user pairs interact on the \wechat{} platform and find useful interaction features for identifying relationship types. 
And then we summarize two major challenges in the edge classification task: The data sparsity problem~\cite{vldb20:Cai} and the computation bottleneck in networks with billions of nodes~\cite{tkde15:Zhang}. 

To address these challenges, we present a Local Community-based Edge Classification (\locec{}) framework that works in three phases, namely \emph{division}, \emph{aggregation} and \emph{combination}.
In the division phase, the entire network is partitioned into ego networks where community detection is performed. 
Such communities are called \emph{local communities} since they only represent the friend circles around an ego node. 
In the aggregation phase, the local communities are classified based on the aggregated user interactions. 
A deep learning model \commcnn{} is proposed to classify local communities into major social connection types.
In the combination phase, the classification results of local communities in different ego networks are unified to form a final relationship type for each edge in the entire network. 

\begin{figure}[!t]
  \centering
  \includegraphics[scale=.35]{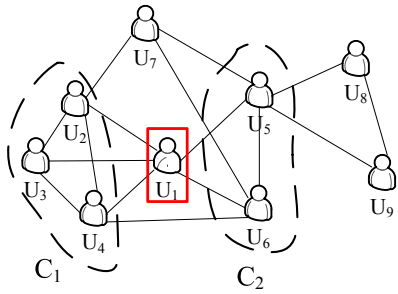}
  \caption{Example Network.}
  \label{fig:eg1}
  \FigBottomVSpace
\end{figure}

\autoref{fig:eg1} illustrates an intuitive example. 
By extracting the ego network of $U_1$, we have nodes $U_2$, $U_3$, $U_4$, $U_5$ and $U_6$ as well as the edges among them.  
Note that the ego node $U_1$ is excluded in its own ego network.
Next, we detect the local communities $C_1$ and $C_2$. 
Afterwards, we classify the edges between the ego node and his friends based on the class label assigned to each local community. 
For example, if we want to identify the relationship type for edge $\langle U_1, U_2 \rangle $, we classify $C_1$ based on the interactions between all user pairs in $C_1$. 
Similarly, we extract $U_2$'s ego network and compute the label for the local community containing $U_1$. 
Then the label of $\langle U_1, U_2 \rangle $  is determined by jointly considering the results from both $U_1$ and $U_2$'s ego networks. 

The contributions of this paper are as follows.

\begin{itemize}
\item We present an in-depth analysis of how different types of user pairs interact on the \wechat{} platform and describe useful features for identifying relationship classes.

\item We propose the \locec{} framework for edge classification. 
\locec{} divides the network into small ego networks for distributed computations and adopts a community-based feature aggregation method to generate dense features for supervised learning tasks.

\item We develop a convolutional neural network-based algorithm, namely \commcnn{}, for effective community classification by exploiting the structural information. 

\item  We perform rigorous experiments to verify the accuracy of \locec{} as well as its usefulness in social advertising. 
Both effectiveness and efficiency are evaluated using \wechat{} network with hundreds of billions of edges. 
To the best of our knowledge, this is the first work that studies edge classification problem at such large scale.
\end{itemize} 

In the remaining, \autoref{sec:background} gives a brief description of the \wechat{} social platform and present an analysis on social relationships. 
We formally define the relationship classification problem and gives an overview of \locec{} in \autoref{sec:problem}. 
The technical details of \locec{} are elaborated in \autoref{sec:methodoloty}. 
Experimental studies are demonstrated in \autoref{sec:experiments}. 
We review the related studies in \autoref{sec:relatedwork} and conclude our paper in \autoref{sec:conclusion}.

\section{Preliminaries}
\label{sec:background}
In this section, we briefly introduce the \wechat{} social platform and conduct an in-depth study of how different types of users interact on \wechat{}.

\subsection{Background}

\wechat{} is a large social mobile application with over one billion monthly active users. 
Users can form mutual friendships by adding each other into their individual contact lists. 
Apart from \emph{instant messaging}, \wechat{} offers a wide range of services such as \emph{group chat} and \emph{moments}. 
A group serve as a virtual chatroom where more than two users can communicate with each other instantly. 
As long as at least one mutual friend exists within a group, users of the same group are not necessarily friends with each other. 
\moments{} are timelines posted by users to share their latest status. 
Friends interact with each other by \emph{liking} or \emph{commenting} on each other's moments. 
Users can create their own timelines or re-post contents posted by others. 
Users can also share mobile games in moments for their friends to join.
Given such social functions of \wechat{}, we next analyze how different types of user pairs interact with each other in multiple ways.

\subsection{Relationship Analysis}

We conducted an analysis on user relationships to figure out (1) what the major relationship types are in the \wechat{} network  and (2) what interaction features best indicate the relationship types.
We first introduce how we obtain ground-truth relationship types and then we describe some findings based on the ground-truth data.

\textbf{Ground-truth data.} 
Users of different ages and genders are paid to participate in an online survey where they provide their chat group names and indicate the true relationship between their contacts. 
Users participated in the survey are fully anonymous. 
In the survey, each mutual user relationship can be \emph{family members}, \emph{colleagues}, \emph{schoolmates} or \emph{others} as the first category, and further subdivided into one of the 13 second categories.  
The detailed design of relationship types is provided in Table~\ref{tab:typell}. 
During the survey, users must specify the first category for the sampled friends, and optionally the second category.
If the second category is not specified, the relationship is marked as unknown.
In total, we have collected $431,409$ relationships with ground-truth labels from $8,805$ users.  
Table~\ref{tab:typell} gives the ratio of each type of relationship. 
As can be seen, the major relationship types, namely family members, colleagues and schoolmates account for $84\%$ of all relationships. 
This indicates that \wechat{} is a social platform mostly for friends who know each other in real life. 
Other than the three major types, friends of similar interests contribute to $9\%$ of edges in the \wechat{} network. 
Due to privacy concern, users left $16\%$ of second category types unspecified. 
In this paper, we only focus on the three major relationship categories, i.e.\ family members, colleagues and schoolmates.

\begin{table}[ht]
	\centering
	\caption{Relationship Types in User Surveys}
	\begin{tabular}{|p{18mm}|p{12mm}|p{13mm}|p{16mm}|}
		\hline 
		First Category & First Ratio & Second Category & Second Ratio \\
		\hline
		\multirow{4}{*}{Family Members} & \multirow{5}{*}{$28\%$} & Next of kin & $0\%$ \\
		&	&		Kin&  $16\%$  \\
		&	& In-law & $5\%$ \\
		&	& Unknown&  $7\%$ \\
		\hline
		\multirow{3}{*}{Colleagues} & \multirow{3}{*}{$41\%$} & Current & $14\%$ \\
		& &		Past &  $25\%$ \\
		& & 	Unknown&  $3\%$ \\
		\hline
		\multirow{5}{*}{Schoolmates} &\multirow{5}{*}{$15\%$} & Primary & $2\%$ \\
		& & 		Middle & $4\%$ \\
		& &		University & $8\%$ \\
		& & Graduate & $0\%$ \\
		& & Unknown & $1\%$ \\
		\hline
		\multirow{5}{*}{Others} &\multirow{5}{*}{$16\%$} &Interest & $9\%$  \\
		& &		Business & $1\%$ \\
		& &      Agent	 &  $1\%$ \\
		&  &      Private & $0\%$ \\
		&  & Unknown  & $5\%$ \\
		\hline 
	\end{tabular}
\label{tab:typell}
\end{table}

\begin{figure}[!t]
  \centering
  \includegraphics[scale=.41]{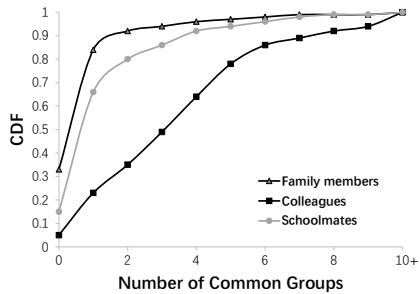}
  \caption{Chat Group Analysis.}
  \label{fig:ana-group}
  \FigBottomVSpace
\end{figure}

\textbf{Chat groups.} 
Next we analyze group chats between different types of user pairs. 
\autoref{fig:ana-group} illustrates the cumulative distribution of number of common groups between friends. 
As can be seen, more than $30\%$ of family member pairs share no common groups. 
Most family members ($>80\%$) share at most one group, while schoolmates have slightly more common groups such that more than $30\%$ of them share at least two groups. 
Colleagues have the largest number of common groups, as more than $40\%$ of colleagues share less than three common groups. 
Group names sometimes indicate relationships between group members. 
For example, \emph{Class X in X Middle school} and \emph{X Department in X Company} imply \emph{schoolmates} and \emph{colleagues} respectively. 
We adopt a rule-based mining method to identify the relationship types of friend pairs within the same group by matching group names with specific patterns, and the result is presented in Table~\ref{table:group}. 
As can be seen, relationship inference based on group name can achieve fairly high precision, e.g.\ above 0.7 for all types.
However, the recall is extremely low, since the majority of groups do not have indicative names and around $20\%$ of friend pairs are not included in any common groups. 
In order to tackle the sparseness of user interaction features such as common group names, we need to find an approach that can generate  features for less interacted user pairs.

\begin{table}[h]
	\centering
	\caption{Group Name Classification Performance} 
	\begin{tabular}{|c|c|c|c|}
		\hline
		Relationship Type & Precision  & Recall & F1-score  \\
		\hline
		Family Members& 0.705 & 0.014 & 0.027   \\
		Colleague & 0.821 & 0.005 & 0.01  \\
		Schoolmates & 0.934 & 0.008 & 0.016 \\ \hline		
	\end{tabular}
	\label{table:group}
\end{table}

\begin{figure}[t]
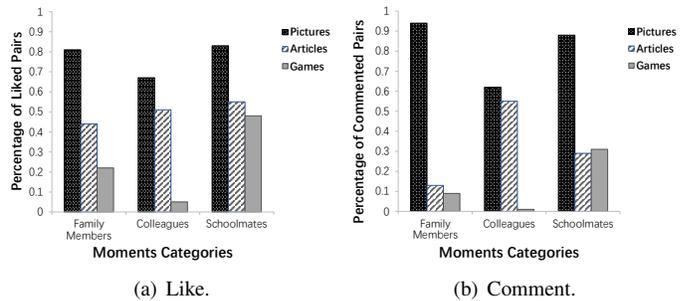

	\begin{center}
		\hspace*{-0.2in}
		\subfigure[Like.]{
			\includegraphics[scale=.35]{\Pic{pdf}{analysis-moments1}}
			\label{fig:moment1}
		}
		\hspace*{-0.11in}	
		\subfigure[Comment.]{
			\includegraphics[scale=.35]{\Pic{pdf}{analysis-moments2}}
			\label{fig:moment2}
		}		
		\caption{Percentage of Interactions under Different Moment Types.}
		\label{fig:moments}
	\end{center}
  \FigBottomVSpace
\end{figure}

\begin{figure}[!t]
  \centering
  \includegraphics[scale=.41]{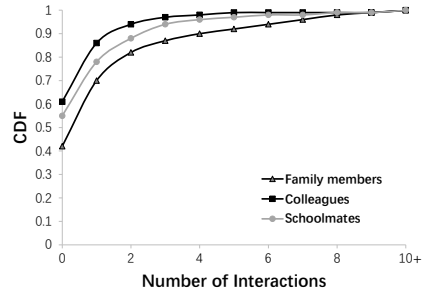}
  \caption{Moments Analysis.}
  \label{fig:moment-cdf}
  \FigBottomVSpace
\end{figure}

\textbf{\moments{}.} 
Next we analyze user interactions in the \moments{} over thirty consecutive days. 
In particular, We investigate the \moments{} posts that fall into the following three categories: \emph{pictures}, \emph{articles} and \emph{games}.
\autoref{fig:moments} shows the ratio of user pairs of different types that have interacted under a particular moment category.
We examine the \emph{like} and \emph{comment} behaviors separately. 
As shown in \autoref{fig:moment1}, different types of user pairs tend to \emph{like} pictures more than articles and games. Colleagues and schoolmates like articles more than family members. 
And schoolmates have the highest probability of liking game posts. 
\autoref{fig:moment2} demonstrates how user pairs of different relationships \emph{comment} on each others' posts. 
As can be seen, all users tend to comment on pictures more than the other two categories. 
Colleagues have a very low appearance of discussing games but a rather high probability of commenting on shared articles. 
Family members and schoolmates mostly comment on pictures while over $30\%$ of schoolmates also talk under a game post.
As demonstrated, different types of user pairs interact differently in the \moments{}, which implies that user interaction features might be useful for relationship classification. 
However, \autoref{fig:moment-cdf} illustrates the cumulative distribution of user interactions and reveals that many user pairs have no interaction in the \moments{} regardless of the relationship types. Due to the sparseness  of potentially useful user interaction features, we are motivated to make use of the limited user interactions to generate  features for most user pairs in the network.

\textbf{Solution discussion.} 
The aforementioned analysis manifests that user interactions can be useful for edge classification. 
However, most user pairs do not have enough interactions or common groups to yield useful features. 
An intuitive approach to address the above issue is to aggregate the features of user pairs that are of the same type to generate denser feature representations. 
Now we analyze how to cluster user pairs of the same type without knowing their relationship types.

\begin{figure}[!t]
  \centering
  \includegraphics[scale=.38]{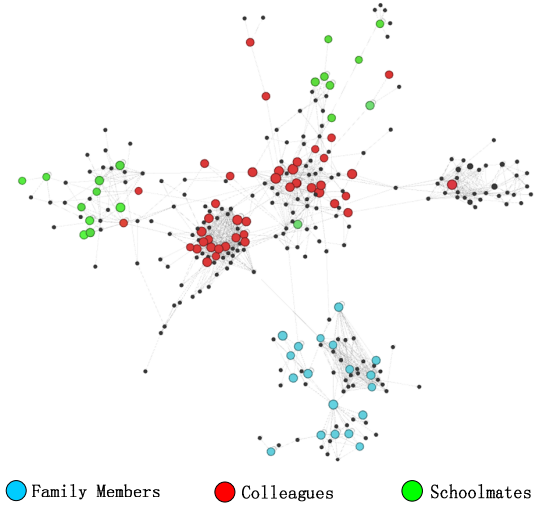}
  \caption{Visualization of Labeled Friends.}
  \label{fig:ana-ego}
  \FigBottomVSpace
\end{figure}

\autoref{fig:ana-ego} illustrates a sample ego network of a surveyed user. 
This network contains the ego user's friends as well as the friendships between them. 
Each color identifies a certain relationship type, and the small black nodes indicates the friends whose types are not specified. 
We can observe two important facts form such visual analysis.
First,  users that are closely connected with each other often have the same relationship type with the ego node. 
Second, friends of the same type could possibly be placed into multiple clusters in the network.

\balance

Based on the above analysis, we know that 
\begin{itemize}
\item \emph{Family members}, \emph{colleagues} and \emph{schoolmates} dominates the relationships on \wechat{}. 

\item \wechat{} offers a wide range of social features like \emph{chat groups} and \emph{moments}. Users' social behaviors could be used as features for classification but the behavioral data is very sparse.

\item Friends that are closely connected often share the same relationship type with each other. This finding could be used for aggregating user features in order to avoid feature sparseness.
\end{itemize}

\begin{figure*}[!t]
  \centering
  \includegraphics[scale=.495]{\Pic{pdf}{framework}}
  \caption{The \locec{} Framework.}
  \label{fig:framework}
  \FigBottomVSpace
\end{figure*}

\vspace*{0.2in}
\section{Problem Definition and Solution Overview}\label{sec:problem}

In this paper, we study the problem of classifying relationships in social networks into predefined social types.

Given a social network $G=(V,E)$ where $V$ is a set of nodes denoting the users and $E$ is a set of edges denoting the user relationships. 
$n=|V|$ and $m=|E|$ are the number of nodes and edges in the network respectively. 
User feature matrix $F = \{f_1, f_2, ... ,f_n\}$ encodes the user features such as \emph{gender}, where the feature vector of a node $v\in V$ is denoted as $f_v$. 
User interactions are represented by $\mathbf{I} = \{I^1,I^2,...,I^{|\mathbf{I}|} \}$, where $I^j$ is an $n \times n$ matrix. For any edge $\langle u,v \rangle\in E$, we have $I^j_{uv}$ indicating the number of times that $u$ and $v$ have interacted on the $j$th dimension. In total, we have $|\mathbf{I}|$ interaction dimensions including  \emph{messaging}, \emph{commenting}, \emph{reposting} or \emph{liking} each other's posts. 
As most of the real-world social networks such as those from \wechat{}, Facebook and LinkedIn are undirected~\cite{xiao:kdd19,backstrom:www11}, 
we presume $G$ to be undirected in the rest of the paper. 

Given  a set of predefined labels $L$, for each edge $e\in E$, there is an underlying label $ l\in L$. 
We presume that each edge can only have one relationship type, and edges with multiple types should be labeled by their principal types\footnote{We leave the multi-type relationship mining as a future work.}. 
Given a small set of edges $E_{labeled}$ with ground truth labels, the relationship classification problem can be formulated as a standard multi-label classification problem that predicts the correct label for all edges in $E \setminus E_{labeled}$.

In order to address the sparsity issue as well as handling the networks with billions of nodes, we design a customized divide-and-combine method and further propose the Local Community-based Edge Classification (\locec{}) framework. 
\autoref{fig:framework} exhibits an overview of the proposed method. 
Specifically, \locec{} comprises a three-phase processing:

\begin{itemize}
\item \textbf{Phase I: Division.} 
This is the \emph{local community detection} phase where ego networks are extracted and community detection is performed. 
For each node $v\in V$, we obtain the ego network $G_v$ of $v$ by extracting the neighbors of $v$ as well as the edges between nodes in $G_v$. 
	
\item \textbf{Phase II: Aggregation.}  
This is the \emph{community classification} phase where user features are gathered together for classifying local communities into predefined relationship types. 
It consists of two stages: feature computation and community classification. 
Given a local community $C$, for each $v\in C$, we aggregate $v$'s interactions with all members in $C$ to obtain $v$'s interaction feature vector $I^C_v$. 
Then we compute a \emph{tightness} value between each node $v\in C$ and local community $C$ to measure how closely this node is connected to the community. 
After that, we arrange the features of the top $k$ nodes with highest tightness values to form a feature matrix that is further fed to a prediction model. 
The output of the community classification is constructed as a softmax vector $r^C$. 
	
\item \textbf{Phase III: Combination.} 
This is the \emph{edge classification} phase where the classification results of local communities are combined to form the final type label for each edge. 
For each edge $\langle u,v \rangle \in E$, we concatenate community classification results $r^{C_u}$ and $r^{C_v}$ with tightness values to train a logistic regression model for predicting the relationship type of edge $\langle u,v \rangle$.
\end{itemize}

\autoref{table:notation} summarizes the frequently used notations in this paper.

\begin{table}[htbp]
	\small
	\centering
	\caption{Summary of Notations}
	\begin{tabular}{|c|c|}
		\hline
		Notation & Meaning\\
		\hline
		$G=(V,E)$  & Graph $G$ with node set $V$ and edge set $E$. \\ \hline
		$G_v$  & Ego network of node $v$. \\ \hline	
		$L$ & The set of relationship type labels. \\ \hline
		$F$ & User feature matrix. User $i$ give row $f_i$. \\ \hline
		$|f|$ & Length of users' individual feature vector. \\ \hline
		$\mathbf{I}$ & User interaction matrices $\mathbf{I} = \langle I^1,...,I^{|\mathbf{I}|} \rangle $.\\ \hline
		$I^i_{uv}$ & Interaction of  $\langle u,v \rangle $ on  dimension $i$.\\ \hline
		$C$ & Local community $C$. \\ \hline
		$r^C$  & Classification result vector for $C$. \\ \hline
		$I^C_u$ & $u$'s interaction features regarding  community $C$. \\ \hline
		$k$ & Number of rows in a feature matrix. \\ \hline
		$f_{\langle u,v \rangle}$ & Feature vector  for edge classification. \\ \hline
	\end{tabular}
	\label{table:notation}
\end{table}

\newpage
\section{Methodology}\label{sec:methodoloty}
In this section, we describe the technical details of each phase in the \locec{} framework.

\subsection{Local Community Detection}
In this phase, the global network $G$ is partitioned into ego network $G_u$ for any $u\in V$. 
We define an ego network of user $u$ as the sub-graph around $u$. 
Formally, $G_u = (V_u, E_u)$ is a sub-graph of $G$ where $V_u \subset V $ contains the \emph{ego} node $u$'s friends and $u \notin V_u$. 
$E_u\in E$ contains the edges between nodes in $V_u$. 
We specifically use \emph{friends} to refer to the users in $G_u$ except the ego user $u$. 

Next, we adopt the Girvan-Newman community detection algorithm (GN)~\cite{girvan:02} to detect local communities in the ego networks. 
Note that the ego node $u$ and its adjacent edges are excluded in its own ego network; otherwise, if the ego node is involved while detecting communities, the whole ego network is likely to be identified as one whole community. 

\begin{figure*}[htbp]
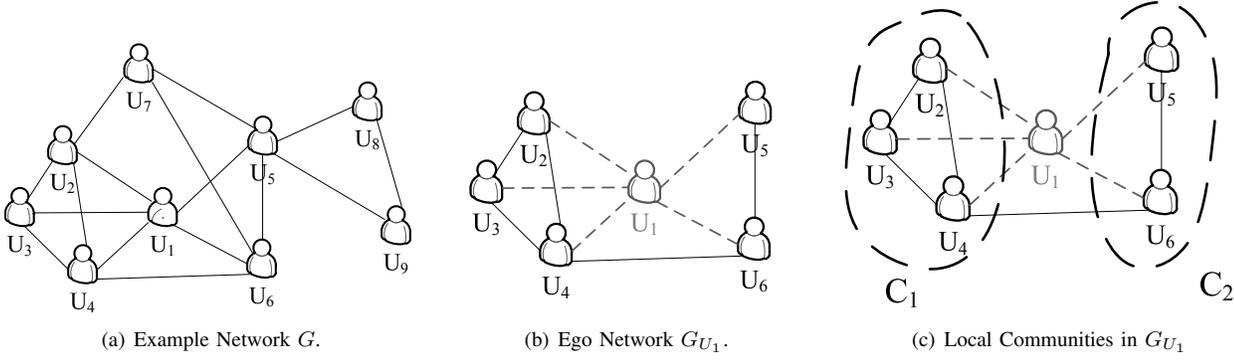

	\centering
	\hspace*{-0.6in}
	\subfigure[Example Network $G$.]{
		\includegraphics[scale=.38]{\Pic{pdf}{example2-0}}
		\label{fig:toy}
	}	
	\hspace*{-0.03in}
	\subfigure[Ego Network $G_{U_1}$.]{
		\includegraphics[scale=.42]{\Pic{pdf}{example2-1}}
		\label{fig:ego}
	}		\hspace*{-0.02in}
	\subfigure[Local Communities in  $G_{U_1}$]{
		\includegraphics[scale=.44]{\Pic{pdf}{example2-2}}
		\label{fig:community}
	} 
	\caption{Illustration of \locec{} \textbf{Phase I: Division} }
	\label{fig:phase_one}
\end{figure*}

Given a network $G$ depicted in \autoref{fig:toy}, \autoref{fig:ego} illustrates the ego network of $U_1$. 
$U_1$ is the ego node and the rest are $U_1$'s \emph{friends}. 
Edges among the ego node's friends are kept in the ego network, whereas edges incident with $U_1$ are dropped (depicted in dashed lines). 
\autoref{fig:community} illustrates the local communities detected in $G_{U_1}$, namely $C_1$ and $C_2$. 

\subsection{Local Community Classification}
Given the local communities in each node's ego network, we train a classification model to label each local community as one of the predefined social types. 
In this phase, a feature matrix is constructed to encode each local community. 
Then the feature matrix is fed into a prediction model for training a classifier. 
We elaborate on the processing in this phase as follows. 


\subsubsection{Feature Matrix Computation}
The community feature matrix consists of feature vectors of users in this community. 
Given a community $C$ and $u\in C$, the user feature consists of two parts. 
The first part $f_u$ contains users' individual attributes such as \emph{gender}. 
The second part $I_u^C$ includes user $u$'s interaction behaviors with other users in $C$ such as \emph{liking} or \emph{commenting} each other's posts. 

For a user $u$, individual features of $u$ are given as a feature vector $f_u$. Note that $f_u$ describes the basic attributes of $u$ and is independent of the local community that $u$ belongs to. The interaction data between two connected users $\langle u,v \rangle$ is given by $\{I^1_{uv}, I^2_{uv},...,I^{|\mathbf{I}|}_{uv}\}$ where $|\mathbf{I}|$ is the number of dimensions that we observe user interactions. We now describe  how we aggregate interaction features between user pairs to form the feature vector $I_u^C$.

Given a friend $u$ in a local community $C$, we  compute the feature value of the $j$th interaction dimension as follows:
\begin{equation} \label{equation:interact}
interact\big(u, C, j\big) = \frac{\sum\limits_{v\in C\backslash u}I^j_{uv}}{\sum\limits_{v,w \in C}I^j_{vw}}
\end{equation}
where $j=1,2,3,\cdots,|\mathbf{I}|$.

Note that even if the interaction between the ego node and a friend may be sparse or empty, Equation~\ref{equation:interact} aggregates the interactions between this friend and all members in the local community to compute his feature value. The $interact\big(u, C, j\big)$ function evaluates how this friend interacts with all members in the same local community.

Then, the interaction feature vector for node $u$ with respect to local community $C$ is given as 
\begin{equation}\label{equation:I} 
	I^C_u = [interact\big(u, C, 1\big),...,interact\big(u, C, |\mathbf{I}|\big)]
\end{equation}

To arrange the feature vectors in a matrix, members in local communities need to be placed in an ordered sequence of fixed length. We define \emph{tightness} values between each node $v$ and its belonged local community $C$ to sort the community members. The intuition is, nodes that are connected with more nodes in the same local community and fewer nodes in other local communities should play more important roles in representing this community and have higher tightness values. 

Formally, given the ego node $v$ and his friend $u$ in a local community $C$, let $G_v$ be the ego network of $v$ and $C$ be the local community containing $u$, the tightness of a node $u \in C$ is defined as 
\begin{equation}
\label{equation:tightness}
\small 
tightness\big(u,C\big) =  \begin{cases}
1 & |C| = 1\\
\frac{|friend\big(u,C\big)|}{|friend\big(u,G_v\big)|}\times \frac{|friend\big(u,C\big)|}{|C|-1} & |C| > 1\\
\end{cases}
\end{equation}
\noindent where $|C|$ is the number of nodes in community $C$ and $friend\big(u,C\big)$ gives the number of friends that $u$ connects with in local community $C$. Similarly, $friend\big(u,G_v\big)$ gives the number of friends that $u$ have in $v$'s ego network. In a special case where only one node is in the local community, its tightness is set to $1$. 

See local community $C_1$ in \autoref{fig:community} as an example.  For $U_2$ and $U_3$, since they are connected with all other nodes in $C_1$ and have zero connection outside $C_1$ except the ego node, their tightness to $C_1$ is computed as\\
\hspace*{0.2in}$tightness\big(U_2,C_1\big) =tightness\big(U_3,C_1\big) = \frac{2}{2}\times \frac{2}{2} = 1$\\
\\
However for $U_4$, its tightness to $C_1$ is \\
\hspace*{0.2in}$tightness\big(U_4,C_1\big) = \frac{2}{2}\times\frac{2}{3} = 0.67$ \\
\\
since it is connected with $U_6$ that is not in $C_1$.

Next, we take the user features as well as interaction features  of the top $k$ users with the highest tightness values to form a $k \times \big(|\mathbf{I}|+|f|\big)$ matrix as the feature matrix of a local community. This feature matrix serves as the  feature representation for each local community in subsequent classification tasks. 

Algorithm~\ref{algo:phase2-1} gives the details of feature matrix construction.
For each node in a local community, lines 3 to 5 compute its feature vector while lines 6 and 7 decide its position in the final feature matrix.
Lines 9 to 11 takes the top $k$ users with the highest tightness values in the local community to form the final feature matrix of the community.
Algorithm~\ref{algo:phase2-1} is repeated for each local community in each ego network.

\begin{algorithm}[h]
	\caption{\textbf{Feature Matrix Construction}}
	\label{algo:phase2-1}
	\SetKwInOut {Input}{input}
	\SetKwInOut {Output}{output}
	\SetKwProg{Function}{Function}{}{}
	\Input {1. Local community $C$\\
		2. User feature matrix $F$\\
		3. User interaction matrices $\mathbf{I}$
	}
	\Output {Local community $C$'s feature matrix}
	\vspace*{0.1in}
	Initialize maximum heap $Q = \emptyset$;\\
	\ForEach{$u\in C$}{
		\For{$j=1,2,3,\cdots,|\mathbf{I}|$}{Compute $interact(u,C,j)$ with Equation~\ref{equation:interact};\\}
		Compute $I_u^C$ with Equation~\ref{equation:I};\\
		Compute $tightness(u,C)$ with Equation~\ref{equation:tightness};\\	
		Insert $u$ into $Q$ with value $tightness(u,C)$
	}
	Initialize $M$ to an empty matrix;\\
	\For{$i=1,\cdots,k$}{
		Pop $u$ with the largest value in $Q$;\\
		Append $[I_u^C, f_u]$ to $M$;
	}
	Return $M$;
\end{algorithm}

\subsubsection{Community Classification Model}\label{sec:commcnn}

We now discuss the choices of classification models given the feature matrix representation of local communities. 

\textbf{\xgb{}.} \xgb{}~\cite{chen:kdd16}  is a popular gradient boosting framework for solving industrial machine learning problems. Since \xgb{} takes feature vectors instead of matrices, we compute the mean and standard deviation of each feature dimension regarding all nodes in a local community to form the feature vector of a community. \xgb{} decides the class label for each local community.

Even though \xgb{} is simple and effective,  taking the mean and deviation reduces the information contained in the feature matrix. Hence, we seek to design a more comprehensive model for community classification.

\textbf{\commcnn{}.}
Graph Convolutional Networks (GCN)~\cite{will:17nips,wu:19gcn} have recently been proposed to handle embedding tasks on network data. 
GCN allows features on each node to be propagated via convolutional kernels to generate embedding for individual nodes as well as graphs. 
However, since each community has its  own structure, GCN requires a training process for each community which is infeasible in our circumstance. Next, we develop a convolutional neural networks-based prediction model named \commcnn{} for community classification. 
In the following, we introduce the details of \commcnn{}.
 
\commcnn{} takes a feature matrix as input and matrices with less than $k$ rows are padded with zeros. We process the matrix with three types of convolutional kernels.
The first convolutional kernel is a popular $3\times 3$ \emph{square kernel}  where  adjacent rows and columns are considered. This kernel design is widely adopted in computer vision problems such as object detection~\cite{girshick:iccv15}. However, unlike image matrices where feature positions are fixed, our feature columns are ordered randomly. Hence we introduce two additional kernels, namely the \emph{wide kernel} and the \emph{long kernel}:

The wide kernel is of size $1\times (|\mathbf{I}|+|f|)$ that looks at all features of the same node as a whole such that all feature attributes of the same user can be compared together. 
The long kernel is of size  $k\times 1$ in order to compare the values of all nodes in each feature dimension.

The design of \commcnn{} is given in \autoref{fig:cnn}.  We have three types of convolution kernels, namely \emph{square}, \emph{wide} and \emph{long}. Wide and long convolutions are followed by a $1\times 1$ convolution and a \emph{global max pooling} layer while the square convolution layer is followed by two \emph{Square Convolution Modules}, consisting of  a $3\times 3$ convolution layer and a max pooling layer,  to abstract the features to a deeper level. 
To summarize, the \emph{Convolution Model} in \autoref{fig:cnn} consists 3 layers in  \emph{wide} and \emph{long} convolutions and 7 layers in  \emph{square} convolutions. The output of Convolution Model is sent to two fully connected layers and a softmax function at the end. 

The output of \commcnn{} model is a vector of length $L$ where   $L$ is the set of relationship types. Each dimension in the output vector gives the probability of the input community belonging to a particular relationship type.

\begin{figure}[t]
  \centering
  \includegraphics[scale=.38]{\Pic{pdf}{cnn}}
  \caption{\commcnn{} Model.}
  \label{fig:cnn}
  \FigBottomVSpace
\end{figure}


\subsection{Edge Labeling}
Given the classification results of local communities, we now proceed to decide the relationship types of all edges. 
For an edge $\langle u,v \rangle  \in E$, we  look at the local community that $u$ belongs to in $v$'s ego network as well as the local community that $v$ resides in $u$'s ego network. Intuitively, if these two local communities are classified as the same social  type, then $\langle u,v \rangle $ is labeled as this particular relationship type. 

\begin{figure}[h]
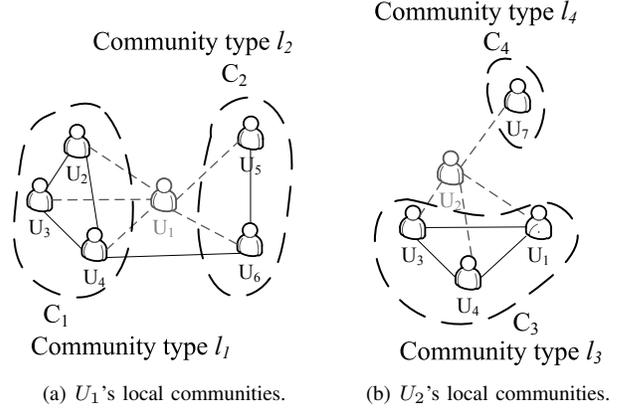

	\begin{center}
		\hspace*{-0.1in}
		\subfigure[$U_1$'s local communities.]{
			\includegraphics[scale=.33]{\Pic{pdf}{example4-1}}
			\label{fig:4-1}
		}	
		\hspace*{-0.08in}
		\subfigure[$U_2$'s local communities.]{
			\includegraphics[scale=.335]{\Pic{pdf}{example4-2}}
			\label{fig:4-2}
		}		
		\caption{Local communities of $U_1$ and $U_2$ .}
		\label{fig:4}
	\end{center}
  \FigBottomVSpace
\end{figure}

Consider our running example in  \autoref{fig:eg1},  \autoref{fig:4} shows the local communities in $U_1$ and $U_2$'s ego networks. Assume that we wish to decide the relationship label of edge $\langle U_1, U_2 \rangle $. We compare the social type of $C_1$ and $C_3$, which are $l_1$ and $l_3$. If $ l_1 = l_3 = l_{same}$, we label $\langle U_1, U_2 \rangle $ as $l_{same}$.

However, in the above example,  $C_1$ and $C_3$ may not be labeled as the same type, \ie $l_1\neq l_3$. To solve such cases, we  design a supervised learning approach to effectively combine the classification results of local communities computed in  individual users' ego networks. 
Specifically, we utilize a  logistic regression (LR) model~\cite{peng:02} as a multi-label classifier to predict the edge label for each edge. 

We now describe how to compute the features for each edge. 
For an edge $\langle u,v \rangle $, assuming $C_u$ is the local community in $v$'s ego network that $u$ resides in and  $C_v$ is the local community in $u$'s ego network that $v$ resides in.

For \xgb{}, classification result vectors $r^{C_u}$ and $r^{C_v}$ are constructed by concatenating the values  of the leaf nodes on the final layers of generated trees for classifying  $C_u$ and $C_v$ respectively~\cite{he:2014adkdd}.

For \commcnn{},  given an edge $\langle u,v \rangle $, $P(C_u, l)$ gives probability of $C_u$ being classified as type $l$ for $l\in L$ and $P(C_v,l)$ gives the probability of $C_v$ being classified as type $l$. $L$ is the set of social labels. We have
$r^{C_u} = [P(C_u, l)~~ \forall l\in L] $ and $r^{C_v} = [P(C_v, l)~~ \forall l\in L] $.

In addition, we also need to consider how close $u$ and $v$ belong to their own local communities in order to decide the relationship type on edge $\langle u,v \rangle $. Hence, we include the \emph{tightness} values of each node in his corresponding  community as well as the community classification results produced by in Phase II to form the feature vector of an edge $\langle u,v \rangle $:
\begin{equation}
\label{equation:f}
f_{\langle u,v \rangle } = [tightness(u,C_u), tightness(v,C_v), r^{C_u},r^{C_v}]
\end{equation} where the tightness function is defined in Equation  \ref{equation:tightness}.

Given the feature vectors, we train a logistic regression model to predict the relationship type of each edge. Recall that edge features are rather sparse in the original dataset. After the divide-aggregate-combine process in previous phases, each edge is guaranteed to have a dense feature vector that can be used for classification.
\balance
\newpage

Algorithm~\ref{algo:all} gives the overall computation procedure of \locec{}. Lines 1 to 4 \emph{divide} the entire network into small ego networks and line 5 computes the classification result for local communities detected in ego networks.
In order to decide the relationship type for each edge, lines 6 and 7 \emph{combine} the classification result for local communities as well as tightness values to form a feature vector that is further fed to a classifier for relationship type  prediction. Line 8 returns the predicted label for each edge.

\begin{algorithm}[t]
	\caption{ \textbf{The \locec{} Algorithm}}
	\label{algo:all}
	\SetKwInOut {Input}{input}
	\SetKwInOut {Output}{output}
	\SetKwProg{Function}{Function}{}{}
	\Input {1. Network  $G=(V,E)$\\
		2. User feature matrix $F$\\
		3. User interaction matrices $\mathbf{I}$
	}
	\Output {Relationship type for each edge}
	\vspace*{0.1in}
	\ForEach{$v\in V$}{Compute set of local communities $\mathbf{C}$;\\
		\ForEach{$C\in \mathbf{C}$}{Compute feature matrix for $C$ with Algorithm~\ref{algo:phase2-1};\\
			Predict $C$'s classification result with \xgb{} or \commcnn{};\\	
		}
	}
	\ForEach{$\langle u,v \rangle \in E$}{ Compute $f_{\langle u,v \rangle }$ with Equation~\ref{equation:f};\\
		Predict relationship type using LR algorithm;\\
		Return predicted relationship type for $\langle u,v \rangle $;
	}
	
\end{algorithm}


\section{Experiments}\label{sec:experiments}

In this section, we conduct experimental studies to verify the effectiveness and efficiency of \locec{} on the \wechat{} network. 
All experiments are carried out on 50 Linux servers with 2 Intel(R) Xeon(R) CPU E5-2620 v3 @ 2.40GHz CPU and 128GB RAM.

\textbf{Datasets.} 
The ground truth datasets are collected via user surveys as introduced in \autoref{sec:background}.
We test our methods with ground-truth data as well as the \wechat{} network with hundreds of billions of edges. 
Individual features are extracted from   users' public profiles such as \emph{gender} etc. 
Interaction features include users' social interactions with friends on \moments{}, such as \emph{liking} or \emph{commenting} each other's public posts. 

\textbf{Comparative Methods.} 
We compare the following  methods in our experiments:
\begin{itemize} 
\item \probwp{}~\cite{aggarwal:icde16}. 
This is a label propagation algorithm that utilizes min-hash functions to compute edge similarity for propagation. 
The number of min-hash functions is set to $20$ as indicated in \cite{aggarwal:icde16}. 

\item \economix{}~\cite{aggarwal:icde17}. 
This approach adopts a matrix-factorization algorithm to utilize both textual and structural information. 
In \cite{aggarwal:icde17}, textual content refers to the communication history between two users. 
We consider each interaction together with the number of interaction times as a \emph{word} in order to apply the \economix{} model, since users' communication histories cannot be obtained.  

\item \xgb{}~\cite{chen:kdd16}. 
We train a gradient boosted decision tree (GBDT) model to classify the edges. 
The input feature consists of the individual features of two end users and the interaction feature between them. 

\item \locecxgb{}. 
This is the proposed \locec{} framework using \xgb{} for community classification. 
We use a statistical method to aggregate the features of all nodes in a local community to generate the feature vector for training. 
Specifically, we compute the mean and standard deviation of each feature dimension regarding all nodes in a local community to form the feature vector of a community. 
Values of the leaf nodes are considered as community embedding for predicting edge labels.

\item \loceccnn{}. 
This is the proposed \locec{} framework using \commcnn{} for community classification. 
The \commcnn{} model has three types of convolution kernels to abstract the feature matrix and train a classifier to decide the type of each community. 
Then the edge label is determined with a logistic regression model taking into account the type distribution of local communities as well as tightness values of the end nodes.
\end{itemize}

\textbf{Evaluation Metrics.} 
As F1-score~\cite{goutte:ecir05} is widely adopted in literature to evaluate multi-label classification techniques, we employ the same metric in our experiments. 
We also report the effectiveness of utilizing the edge classification results for social advertising.

\subsection{Parameter Study}

Recall that, for \commcnn{}, a parameter $k$ is used to decide the number of users whose features are used to construct  the feature matrix. 
For local communities with more than $k$ users, we consider the top $k$ nodes with highest \emph{tightness} values. 
For local communities with less than $k$ users, we adopt zero-padding to fill up the spare rows.

\begin{figure}[t]
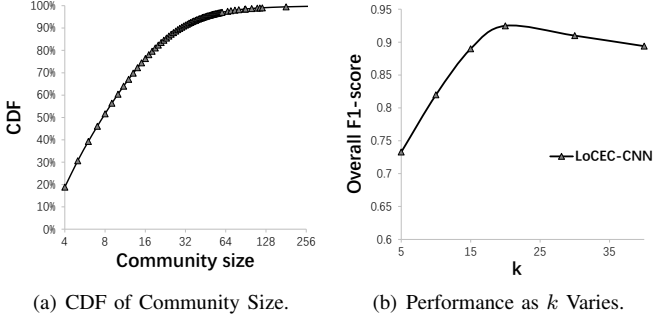

	\begin{center}
		\hspace*{-0.15in}
		\subfigure[CDF of Community Size.]{
			\includegraphics[scale=.45]{\Exp{pdf}{cdf}}
			\label{fig:cdf}
		}		\hspace*{-0.1in}	
		\subfigure[Performance as $k$ Varies.]{
			\includegraphics[scale=.445]{\Exp{pdf}{k}}
			\label{fig:k}
		}
	\end{center}
\caption{Parameter Study.}
\end{figure}

To determine the optimal choice of $k$, we first look into the cumulative distribution of all the local community sizes, as shown in \autoref{fig:cdf}. 
As can be seen, the median size of all communities is $8$. 
Around $80\%$ of all local communities have no more than $20$ users and $90\%$ of the local communities have less than $30$ users. 
We further conduct a preliminary study of $k$ by varying $k$ from $5$ to $40$. 

As can be seen in \autoref{fig:k}, when $k$ is small, the overall F1-score is low since there is not enough information for  \loceccnn{} to learn. 
The performance peaks at $k=20$ and then drops as $k$ increases. 
This is because when $k$ is too big, many feature matrices need to be padded with many zero rows, incurring unwanted noise to the model. 
Hence, in all the experiments, we set $k=20$ for \loceccnn{}. 
On the other hand, \locecxgb{} does not depend on $k$, since it uses mean and deviation to aggregate user features. 
And the baseline methods do not utilize community structure and thus are irrelevant of $k$.

\subsection{Relationship Classification}

Next, we compare proposed methods in terms of the accuracy of edge classification. 

\textbf{Effectiveness study.}
In total, $E_{labeled}$ covers $431,409$ edges, which account for around $0.02\%$ of edges in \wechat{}. 
To compare with the label propagation algorithms, we extract a sub-graph by extending from surveyed users to their friends. We ensure around $40\%$ of edges are given ground truth labels within the sub-graph. 
Note that the identities of surveyed users' friends are fully anonymized.
In particular, this sub-graph contains $42,078$ nodes and $1.1$ million edges.
In this set of experiments, we split the entire dataset into two distinct parts such that $80\%$ of the labeled edges serve as the training set and the other $20\%$ ones are used as the test set. 

\begin{table}[htp]
	\centering
	\caption{Relationship Classification Performance} 
	\begin{tabular}{|c|c|c|c|c|}
		\hline
		Algorithm & Community Type &   Precision  & Recall & F1-score \\
		\hline
        \multirow{4}{*}{\probwp{}}	&	Colleague & 0.804 & 0.778 & 0.791   \\ 
		 & Family Members& 0.787 & 0.776 & 0.782 \\ 
 	& 		Schoolmates & 0.820 & 0.803 & 0.811\\  \cline{2-5}
	& 		Overall & 0.805 & 0.782 & 0.793\\ 
		\hline
		\multirow{4}{*}{\economix{}}	&	Colleague & 0.744 & 0.736 & 0.740   \\ 
		& Family Members& 0.702 & 0.767 & 0.733 \\ 
		& 		Schoolmates & 0.767 & 0.806 & 0.786\\  \cline{2-5}
		& 		Overall & 0.747 & 0.762 & 0.754\\ 
		\hline
		\multirow{4}{*}{\xgb{}}	&	Colleague & 0.714 & 0.673 & 0.693   \\ 
		& Family Members& 0.747 & 0.615 & 0.675 \\ 
		& 		Schoolmates & 0.692 & 0.613 & 0.650\\  \cline{2-5}
		& 		Overall & 0.720 & 0.633 & 0.674\\ 
		\hline
		\multirow{4}{*}{\locecxgb{}}	&	Colleague &  0.754 & 0.862 & 0.804   \\ 
		& Family Members&  0.870 & 0.875 & 0.872 \\ 
		& 		Schoolmates & 0.852 & 0.866 & 0.859\\  \cline{2-5}
		& 		Overall & 0.830 & 0.871 & 0.850\\ 
		\hline
		\multirow{4}{*}{\loceccnn{}}	&	Colleague & 0.866 & 0.875 & 0.870   \\ 
		& Family Members& 0.935 & 0.952 & 0.943 \\ 
		& 		Schoolmates & 0.890 & 0.933 & 0.911 \\  \cline{2-5}
		& 		Overall &  \textbf{0.902} & \textbf{0.931} & \textbf{0.916}\\ 
		\hline
	\end{tabular}
	\label{table:edge}
\end{table}

\begin{figure*}[t]
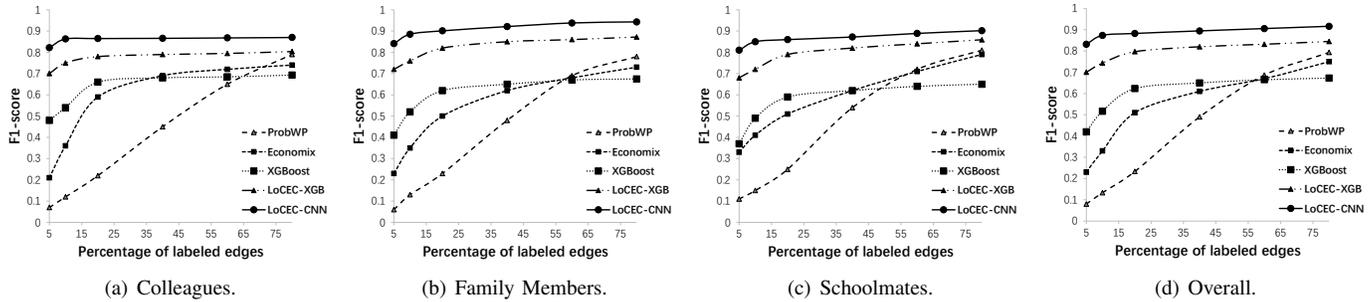

	\begin{center}
		\hspace*{-0.2in}
		\subfigure[Colleagues.]{
			\includegraphics[scale=.34]{\Exp{pdf}{lp1}}
			\label{fig:lp1}
		}		\hspace*{-0.1in}	
		\subfigure[Family Members.]{
			\includegraphics[scale=.34]{\Exp{pdf}{lp2}}
			\label{fig:lp2}
		}		\hspace*{-0.1in}
		\subfigure[Schoolmates.]{
			\includegraphics[scale=.34]{\Exp{pdf}{lp3}}
			\label{fig:lp3}
		}		\hspace*{-0.1in}
		\subfigure[Overall.]{
			\includegraphics[scale=.34]{\Exp{pdf}{lp4}}
			\label{fig:lp4}
		}		
		\caption{Performance of Edge Classification.}
		\label{fig:lp}
	\end{center}
  \FigBottomVSpace
\end{figure*}

\autoref{table:edge} shows the detailed results of relationship classification. 
\probwp{} utilizes a propagation-based approach for edge labeling and achieves an F1-score of $0.793$. 
Based on our analysis in \autoref{sec:background}, in \wechat{}, closely connected users are often linked by edges of the same relationship type, making \probwp{} fairly effective by only considering structural information.
\xgb{} directly uses user features to train a classifier and has the lowest F1-scores.
This is because many user pairs lack interaction data, resulting in bad performance in the recall. 
\economix{} performs better than \xgb{} as it also utilizes structural information for propagating edge labels.
Both \locecxgb{} and \loceccnn{} use the local community structure to aggregate features for predicting edge labels to avoid the sparsity problem.
\loceccnn{} performs the best with  $0.916$ in F1-score and \locecxgb{} is a close runner-up with $0.907$ in F1-score. 
This confirms that the \commcnn{} model is more effective in aggregating features than taking means and deviations.



\textbf{Varying ground truth percentage.}
In the previous experiment,  a sub-graph with $40\%$ of edges are given ground-truth labels.  In this set of experiments, we vary  the percentage of labeled edges from $5\%$ to $80\%$ (out of the $40\%$ of labeled edges) and report the performance of different algorithms in deciding the rest of the edges' types. Note that we only evaluate the labels predicted for edges whose ground truth types are known. \autoref{fig:lp} gives the results.

As can be seen, \loceccnn{} constantly outperforms all baseline methods in terms of F1-score. Runner up algorithm is \locecxgb{} since it also utilizes local community to aggregate user features. When the percentage of  ground truth labels is $5\%$, label propagation algorithms \probwp{} performs poorly with an F1-score less than $0.1$,   while supervised learning algorithms perform much better.
As more labeled edges are given, the performance of \probwp{} increases significantly.  From the user surveys, only $0.02\%$ of edge labels are obtained in the \wechat{} network, label propagation algorithms are not applicable.
\economix{}'s performance also increases as more labels are given since it also utilizes structural information for propagating labels. When a small percentage of edge labels are given, \economix{} gives better results by considering user interactions.
Note that \xgb{} algorithm performs  better than label propagation algorithm when only $5\%$ to $30\%$ of ground truth labels are given. However, \probwp{} and \economix{} achieve higher accuracy  than \xgb{} when $80\%$ of edge labels are visible. This is because, \probwp{} relies heavily on given labels in determining labels for new edges while \economix{} uses matrix factorization to propagate labels to similar nodes.  On the other hand, \xgb{} suffers from the feature sparsity problem which cannot be solved by adding samples.

\subsection{Community Classification Performance}

We now examine the performance of \locecxgb{} and \loceccnn{} in classifying local communities. Note that label propagation-based algorithms  \probwp{} and  \economix{} as well as \xgb{} are  omitted here since they mainly focus on classifying edges  instead of communities. In this experiment, we extract the ego networks of all surveyed users and detect local communities using Girvan-Newman \cite{girvan:02} algorithm. Then, the ground-truth label of a community is determined by the majority type of friends with ground-truth relationship classes. We have around $1.16$ thousand local communities with ground-truth labels, indicating they are \emph{colleagues}, \emph{family members} or \emph{schoolmates}. 
We split  $80\%$ of these labeled communities into training set and $20\%$ into test set.

\begin{table}[htp]
	\centering
	\caption{Relationship Classification Performance} 
	\begin{tabular}{|c|c|c|c|c|}
		\hline
		Algorithm & Community Type &   Precision  & Recall & F1-score \\
		\hline
		\multirow{4}{*}{\locecxgb{}}	&	Colleague & 0.808 & 0.887 & 0.845    \\ 
		& Family Members & 0.892 & 0.914 & 0.903 \\ 
		& 		Schoolmates & 0.884 & 0.889 & 0.886\\  \cline{2-5}
		& 		Overall & 0.863 & 0.900 &0.882\\ 
		\hline
		\multirow{4}{*}{\loceccnn{}}	&	Colleague & 0.879 & 0.903 & 0.891  \\ 
		& Family Members& 0.941 & 0.967 & 0.953 \\ 
		& 		Schoolmates &  0.901 & 0.944 & 0.920 \\  \cline{2-5}
		& 		Overall &  \textbf{0.914} & \textbf{0.940} & \textbf{0.927}\\ 
		\hline
	\end{tabular}
	\label{table:com}
\end{table}

\autoref{table:com} gives the evaluation results of local community classification. As can be seen in Table \ref{table:com}, \loceccnn{} gives the better overall performance with $0.927$ in F1-score. It improves \locecxgb{} by $5.1\%$ with a transformed CNN model. 
Note that the F1-scores of edge classification are slightly lower than those of community classification. We believe this is due to the impurity of community detection results. 
Consider an example scenario where all colleagues connect with a tour guide in a team building trip.
The tour guide is placed in a local community where most of its members are \emph{colleagues} of each other. Hence this particular local community will be labeled as \emph{colleagues} in the ground truth. When \locecxgb{}  and \loceccnn{} classify this particular local community as \emph{colleagues}, it is a true positive. However, when the algorithms try to label the edges between the tour guide and the colleagues, it is hard for the classifier to make a correct decision.
Hence, we argue it is also important to distinguish the impurity in detected local communities and we leave this as future work.

\subsection{Scalability Study}
We now discuss the scalability of our proposed frameworks. Since \loceccnn{} is more effective than \locecxgb{} and they share the same computation framework,  we only apply \loceccnn{} on the entire \wechat{} network for scalability study.
\begin{table}[htp]
	\centering
	\caption{Running Time (hours) of \loceccnn{}} 
	\begin{tabular}{|*{6}{c|}}
		\hline 
		Method & Training & Phase I  & Phase II & Phase III  & Total \\
		\hline
		\loceccnn{} & 4.5 & 46.5 & 15.3 & 7.4 & 73.7 \\
		\hline
	\end{tabular}
	\label{table:runtime}
\end{table}

\autoref{table:runtime} gives the  running hours of \loceccnn{}  taking the entire \wechat{} network as input.
Note that  \commcnn{} model is trained beforehand and it takes 4.5 hours.
We report the averaged results over three runs using $100$ Linux servers with the same \commcnn{}  model. 
Phase I consumes the highest amount of time (around two days) to detect local communities while Phase II together with Phase III take one day to finish for community and edge classification. 

\begin{figure}[t]
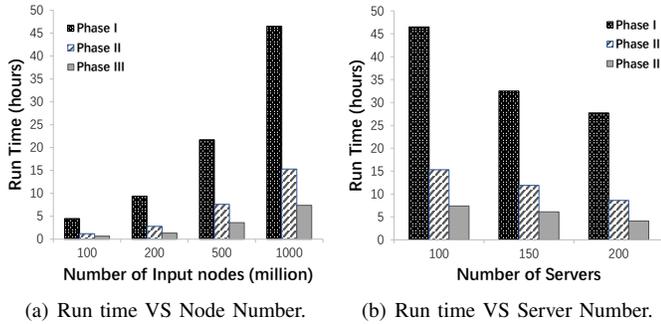

	\begin{center}
		\hspace*{-0.15in}
		\subfigure[Run time VS Node Number.]{
			\includegraphics[scale=.4]{\Exp{pdf}{runtime}}
			\label{fig:runtime}
		}		\hspace*{-0.1in}	
		\subfigure[Run time VS Server Number.]{
			\includegraphics[scale=.4]{\Exp{pdf}{machine}}
			\label{fig:machine}
		}
	\end{center}
\caption{Scalability Study.}
\label{fig:scalability}
\FigBottomVSpace
\end{figure}

\autoref{fig:scalability} analyzes the scalability of \loceccnn{} given different input sizes and computation resources. 
 \autoref{fig:runtime} gives the running hours of the different phases when we increase the number of input nodes from 100 million to 1 billion. As can be seen, the running time increases linearly as the number of input nodes grows. 
 \autoref{fig:machine} shows the running time for \loceccnn{} to process the entire \wechat{} network by  varying  the computation resources. 
As demonstrated, when the number of servers used for computation increases, the computation time decreases accordingly.
This is because, in the \locec{} framework, each node is parsed separately in a streaming scheme in all three phases. This set of experiments show that \locec{} is highly scalable and capable of processing large real world social networks.

\subsection{Application of Edge Classification Results}
\begin{figure}[t]
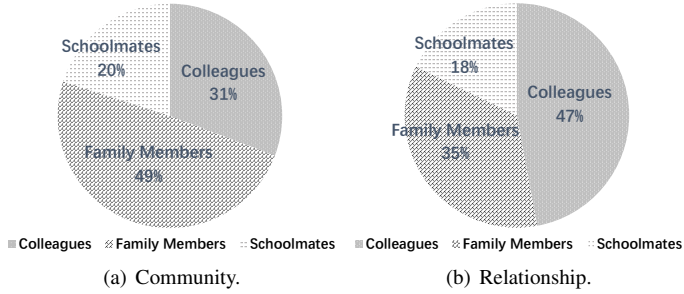

	\begin{center}
		\hspace*{-0.2in}
		\subfigure[Community.]{
			\includegraphics[scale=.23]{\Exp{pdf}{distribution1}}
			\label{fig:distribution1}
		}
	\hspace*{-0.1in}	
		\subfigure[Relationship.]{
			\includegraphics[scale=.23]{\Exp{pdf}{distribution2}}
			\label{fig:distribution2}
		}		
		\caption{Distribution of Community and Relationship Types.}
		\label{fig:distribution}
	\end{center}
  \FigBottomVSpace
\end{figure}

In total, \loceccnn{} classified $5.2$ billion local communities as well as $140$ billion edges  into predefined social types.
The distribution of predicted community types and relationship types are given in \autoref{fig:distribution}.
As can be seen, \emph{family members} and \emph{colleagues} take up the majority of our social circles with \emph{schoolmates} account for around $20\%$ of our friends on \wechat{}.
We notice in the community type distribution, the percentage of \emph{family members} and \emph{colleagues} are  $49\%$ and $31\%$ respectively. However in the relationship type distribution, \emph{family members} and \emph{colleagues} account for $35\%$ and $47\%$ respectively. This is because, the size of family communities are much smaller than that of colleague communities. This is unanimous with our intuition that the social circle in the working environment is often bigger than our families.

\begin{figure}[t]
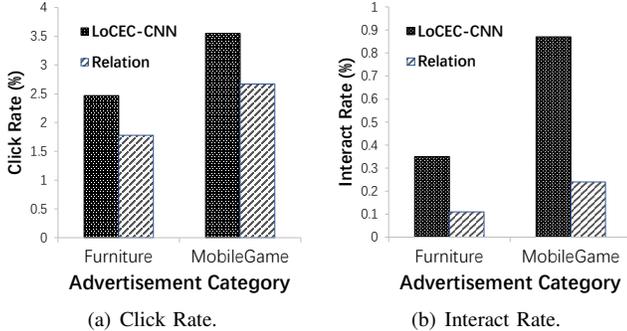

	\begin{center}
		\hspace*{-0.3in}
		\subfigure[Click Rate.]{
			\includegraphics[scale=.42]{\Exp{pdf}{ad1}}
			\label{fig:ad1}
		}	
	\hspace*{-0.05in}
		\subfigure[Interact Rate.]{
			\includegraphics[scale=.42]{\Exp{pdf}{ad2}}
			\label{fig:ad2}
		}		
		\caption{Performance in Social Advertising.}
		\label{fig:ad}
	\end{center}
  \FigBottomVSpace
\end{figure}

The results of \loceccnn{} are deployed in many applications in \wechat{} such as social advertising in \emph{Moments}. 
Advertisements in \wechat{} \emph{Moments} are displayed the same way as users' timelines. Friends can see each other's \emph{like} or \emph{comment} under the same advertisement. They can also \emph{interact} with other friends by replying to their friends' comments. 
Ad owners provide a  set of \emph{seed users} that are known to be interested in the product. 

Here we compare two approaches. Relation method chooses the friends of seed users  while \loceccnn{} selects friends of a specific type.  The same click-through-rate (CTR) scoring function is adopted by both Relation and \loceccnn{} and the friends with the highest scores are returned.
Note that  \emph{furniture} ads are more popular among \emph{family members} while \emph{mobile game} ads attract more interaction among \emph{schoolmates}.

\autoref{fig:ad} shows the results.
\loceccnn{} reaches higher click rate when compared with simply choosing high-score friends. This is because, intuitively, users tend to pay more attention to the furniture that their family members have liked.
Similarly, schoolmates are more likely to be interested in gaming apps that their friends have talked about.  
The advantage of \loceccnn{} is even more obvious when we consider the interact rate in \autoref{fig:ad2}. Interactions are strong indications of a user being interested in the advertised product and \loceccnn{} boosts interaction rate more than twice of Relation. Clearly, the results generated by \loceccnn{} are highly valuable in social applications.


\section{Related Work}
\label{sec:relatedwork}

While most existing studies on graph classification focus on nodes \cite{bilgic:kdd08, zhou:icml05, zhu:icml03}, there is some recent research interest in classifying relationships in online social networks. 
Some studies work on classifying edges as friends and enemies~\cite{agrawal:ijcai13,leskovec:www10,song:kdd05,yang:sigir12,gupta2017prediction}. 
They analyze network topology and employ link prediction methods to identify edges with positive or negative signs. 
The work in \cite{yang:sigir12} also utilizes a behavior relation interplay model that studies behavioral evidence  to infer social interactions. 
However, our work is not to infer the sentiment of social connections but to identify the underlying social relations.
Other works in \cite{backstrom:cscw14,wang:kdd10,li:www14} focus on a particular domain of relationships. 
Wang~\etal~\cite{wang:kdd10} propose a supervised probabilistic model to classify connections into adviser-advisee relationships in a publication network. Backstrom \etal \cite{backstrom:cscw14} study the dispersion of a pair of users' social circles to identify romantic relationships. Li~\etal~\cite{li:www14} adopt a label propagation algorithm for identifying user attributes as well as employer-employee relationships in a workplace. Tang~\etal~\cite{tang:pkdd11} utilize a semi-supervised approach to infer edge types relying on available data in a particular domain and correlation factors of networks. 
The above mentioned methods require domain-specific insight of the network or relationships since they are tailored for a classification problem under specialized assumptions. However in this paper, we aim to solve the relationship classification problem in a more general setting where all major social relation types could be considered.

The work in \cite{aggarwal:icde16} was the first to formally study the problem of edge classification in networks. They consider a large percentage of relationships with ground-truth labels being available. 
They define a structural similarity measure based on the proportion of neighbors that share the same ground-truth label. 
For an unlabeled edge $\langle u,v \rangle $, they look for the top-$k$ nodes that are most structurally similar to $u$, denoted as $S_u$ and repeat the same step for node $v$, obtaining $S_v$.  Then the dominant class label of \emph{labeled}  edges with one end  in $S_u$ and another end in $S_v$ is assigned as the edge label for $\langle u,v \rangle $. Their approach relies heavily on the number of labeled edges. As shown in the experiments in \cite{aggarwal:icde16}, there is a significant drop in the accuracy when the percentage of unlabeled edges increases from $10\%$ to $30\%$. 
However, in a real-life social network like \wechat{}, the percentage of unlabeled relationship is up to $99\%$, rendering their approach inapplicable. Furthermore, Aggarwal~\etal~\cite{aggarwal:icde17} extend their method to consider text associated with user relationships such as communication history. 
They propose a matrix factorization based algorithm to utilize both textual and structural information by treating each relationship as a document with words. 
In this work, we aim to utilize user interactions instead of communication histories as relationship features and propose a local community-based approach to handle the sparsity problem.

\section{Conclusion}\label{sec:conclusion}

In this paper, we study the problem of relationship classification in large online social networks. 
We carry out an in-depth analysis of user relationships on the \wechat{} platform. 
Based on the findings from the analysis, we propose the \locec{} framework to tackle the sparsity and scalability challenges.
Moreover, a \commcnn{} model is crafted to classify local communities into predefined social types, and a logistic regression model is further adopted to expand the community types onto edges. 
Experimental comparison against state-of-the-art relationship classification techniques confirms \locec{}'s superiority in terms of effectiveness and scalability. 
We believe this is the first work that systematically solves the edge classification problem in a real-world billion-scale network.

\section*{Acknowledgment}


We would like to thank Lingling Yi and Hao Gu for sharing the ground-truth data obtained from user surveys. 
We would also like to thank the anonymous reviewers for their valuable comments and constructive suggestions that helped improve the paper.
Qian Lin is supported by Singapore Ministry of Education Academic Research Fund Tier 3 under MOE's official grant number MOE2017-T3-1-007.

\balance

\newcommand{\RefFile}{references-trim}

\bibliographystyle{IEEEtran}
\bibliography{\RefFile}

\end{document}